\documentclass{aastex62}
\usepackage{multirow}
\newcommand{\cawav}{Ca\,{\sc ii}~8542\,\AA}
\newcommand{\ca}{Ca\,{\sc ii}}
\newcommand{\ha}{H$\alpha$}
\newcommand{\hb}{H$\beta$}
\newcommand{\cmcub}{cm$^{-3}$}
\newcommand{\kms}{km\,s$^{-1}$}
\newcommand{\vt}{v_{\rm tur}}
\newcommand{\pg}{p_{\rm g}}
\newcommand{\nec}{n_{\rm e}({\rm c})}
\newcommand{\nes}{n_{\rm e}({\rm s})}
\defcitealias{Kuridzeetal2019}{KMM}

\received{June 26, 2019}
\revised{September 4, 2019}
\accepted{September 11, 2019}

\shorttitle{Spectral diagnostics of cool flare loops}
\shortauthors{Koza, Kuridze, Heinzel et al.}

\begin{document}
\title{Spectral diagnostics of cool flare loops observed by SST: \\ I. Inversion of the \cawav\ and \hb\ lines}

\correspondingauthor{J. Koza}
\email{koza@astro.sk}

\author[0000-0002-7444-7046]{J\'{u}lius Koza} 
\affil{Astronomical Institute, Slovak Academy of Sciences, 059 60 Tatransk\'{a} Lomnica, Slovakia}

\author[0000-0003-2760-2311]{David Kuridze} 
\affiliation{Institute of Mathematics, Physics and Computer Science, Aberystwyth University, Ceredigion, Cymru, SY23 3BZ, UK}
\affiliation{Astrophysics Research Centre, School of Mathematics and Physics, Queen's University Belfast, Belfast BT7 1NN, UK}
\affiliation{Abastumani Astrophysical Observatory at Ilia State University, 3/5 Cholokashvili Avenue, 0162, Tbilisi, Georgia}

\author{Petr Heinzel}
\affiliation{Astronomical Institute, The Czech Academy of Sciences, 25165 Ond\v{r}ejov, Czech Republic}
 
\author[0000-0001-8489-4037]{Sonja Jej\v{c}i\v{c}} 
\affiliation{Faculty of Education, University of Ljubljana, Kardeljeva plo\v{s}\v{c}ad 16, 1000 Ljubljana, Slovenia} 
\affiliation{Faculty of Mathematics and Physics, University of Ljubljana, Jadranska 19, 1000 Ljubljana, Slovenia}

\author[0000-0002-6547-5838]{Huw Morgan}
\affiliation{Institute of Mathematics, Physics and Computer Science, Aberystwyth University, Ceredigion, Cymru, SY23 3BZ, UK}

\author[0000-0001-9378-7487]{Maciej Zapi\'{o}r}
\affiliation{Astronomical Institute, The Czech Academy of Sciences, 25165 Ond\v{r}ejov, Czech Republic}

\begin{abstract}
Flare loops form an integral part of eruptive events, being detected
in the range of temperatures from X-rays down to cool
chromospheric-like plasmas. While the hot loops are routinely observed
by the Solar Dynamics Observatory's Atmospheric Imaging Assembly
(SDO/AIA), cool loops seen off-limb are rare. In this paper we employ
unique observations of the SOL2017-09-10T16:06 X8.2-class flare which
produced an extended arcade of loops. The Swedish 1-m Solar Telescope
(SST) made a series of spectral images of the cool off-limb loops in
the \cawav\ and the hydrogen \hb\ lines. Our focus is on the loop
apices. Non-LTE spectral inversion is achieved through the
construction of extended grids of models covering a realistic range of
plasma parameters. The Multilevel Accelerated Lambda Iterations (MALI)
code solves the non-LTE radiative-transfer problem in a 1D
externally-illuminated slab, approximating the studied loop
segment. Inversion of the \cawav\ and \hb\ lines yields two similar
solutions, both indicating high electron densities around $2 \times
10^{12}$\,\cmcub\ and relatively large microturbulence around
25\,\kms. These are in reasonable agreement with other independent
studies of the same or similar events. In particular, the high
electron densities in the range $10^{12} - 10^{13}$\,\cmcub\ are
consistent with those derived from the SDO's Helioseismic and Magnetic
Imager white-light observations. The presence of such high densities
in solar eruptive flares supports the loop interpretation of the
optical continuum emission of stars which manifest superflares.
\end{abstract}



\section{Introduction} 

Solar flares are rapid energy releases within active regions caused by
the reconnection of the coronal magnetic field resulting in plasma
heating to temperatures beyond tens of millions of kelvins. A
significant amount of the released energy is transported along the
magnetic loops to the lower solar atmosphere via accelerated
particles, magnetohydrodynamic waves, and thermal conduction
\citep[see Chapter 16]{Hirayama1974,Aschwanden2005}.The bulk of flare
energy dissipates upon reaching the dense loop footpoints. As most of
the flare energy is deposited in the chromosphere, a strong
evaporation of the heated plasma occurs
\citep[e.\,g.,][]{Neupert1968,Fisheretal1985c,Fisheretal1985b,Fisheretal1985a,GrahamCauzzi2015}. A
detailed EUV spectroscopic analysis of explosively evaporating plasma
following an M1.6-class flare by \citet{Gomoryetal2016} found electron
densities up to $3.16 \times 10^{10}$\,\cmcub\ at a temperature of
1.56\,MK. The evaporation provides an injection of plasma into flare
coronal loops (hereafter flare loops) and leads to the formation of
bright high-density structures at the apex of the flare loop arcade.
Later, the evaporated plasma cools and subsequently drains down along
the loops towards the chromosphere in the form of discrete plasma
blobs called ``coronal rain''
\citep{Antolinetal2010,Vashalomidzeetal2015,Jingetal2016}.  In this
context we refer to \citet{Svestka2007} who criticized the widespread
misnomers ``post-flare loops'', ``loop-prominence systems'', or ``loop
prominences''. Observations show that such loop systems are not truly
post-flare phenomena but are an integral part of the flare event and
therefore the terms ``eruptive flare loop system'' or ``flare loops''
are more relevant.  However, quantitative measurements of the
fundamental plasma parameters of flare loops and their bright apices
such as temperature, density, gas pressure or magnetic field strength
are rare.

Since its launch on 2013 June 23, the {\it Interface Region Imaging
  Spectrograph} \citep[IRIS;][]{DePontieuetal2014} brings new
important insights into the structure and dynamics of flare loops
during the evaporation, cooling, and subsequent draining phases. The
study of \citet{Lacatusetal2017} showed remarkably broad and
redshifted emission profiles of Mg\,{\sc ii}, C\,{\sc ii}, and
Si\,{\sc iv} lines observed by IRIS in on-disk flare loops of the
X2.1-class flare. The authors interpret the very broad line profiles
by unresolved Alfv\'{e}nic motions, waves or turbulence whose energy
exceeds the radiation losses in the IRIS Mg\,{\sc ii} lines by an
order of magnitude. Extremely broad Mg\,{\sc ii} emission profiles,
observed in on-disk flare loops, have been also reported in
\citet{Mikulaetal2017} and explained as due to large unresolved
non-thermal motions. Analyzing IRIS observations of on-disk flare
loops of an M-class flare, \citet{Brannon2016} also found significant
non-thermal broadening of the Si\,{\sc iv}\,1402.772\,\AA\ line,
ranging from $\approx 12$\,\kms\ near the loop apices up to $\approx
60$\,\kms\ near the footpoints, and an approximately uniform and
constant electron density of $10^{11}$\,\cmcub\ in the loops.

In September 2017, the Active Region (AR) NOAA 12673 produced a series
of powerful X-class flares as it rotated from the disk center to the
limb \citep{Verma2018,Romanoetal2019}. On 2017 September 10, the AR
was just behind the western limb when it produced the
SOL2017-09-10T16:06 X8.2-class flare, ranked as the second-largest
flare in the solar cycle 24. The flare triggered a very fast coronal
mass ejection \citep{Veronigetal2018} followed by significant space
weather and heliospheric effects including a Solar Energetic Particles
(SEP) event \citep{Kurtetal2018}. Analysis of elemental abundances of
the flare loops was performed in \citet{Doscheketal2018} using spectra
obtained by the Hinode's {\it Extreme-ultraviolet Imaging
  Spectrometer} \citep[EIS;][]{Culhaneetal2007}. In some loops, they
find that the abundances are coronal at the loop apices or cusps,
gradually changing to photospheric towards the loop footpoints (see
Figure~4 therein). Remarkably, using the intensity ratio of the
Ca\,{\sc xiv} and Ar\,{\sc xiv} lines near 194\,\AA\, they found
electron densities of $2-5.8 \times 10^{10}$\,\cmcub\ in the bright
cusp of flare loops. On the other hand, \citet{Jejcicetal2018} found
relatively high electron densities ranging from $10^{12}$ to
$10^{13}$\,\cmcub\ in their analysis of the same flare loops observed
in white-light (WL) continuum by the {\it Solar Dynamics
  Observatory's} \citep[SDO;][]{Pesnelletal2012} {\it Helioseismic and
  Magnetic Imager} \citep[HMI;][]{Schouetal2012,Scherreretal2012}
under the assumption of optically thin flare loop plasma in the
continuum at 6173\,\AA. They concluded that the hydrogen Paschen and
Brackett recombination continua are dominant in cool flare loops ($T
\lesssim 2 \times 10^4$\,K), while the hydrogen free-free continuum
emission is dominant for warmer and hot loops ($T > 2 \times 10^4$)\,K
(Figure~4 therein). Finally, \citet*[hereafter
  \citetalias{Kuridzeetal2019}]{Kuridzeetal2019} presented a unique
measurement of the magnetic field of this flare loop arcade yielding
the field strength as high as 350\,G at heights up to 25\,Mm.

An extensive compilation of earlier high electron-density measurements
in flares is given in \citet{Svestka1972}. More recently, high
electron density of the order of $10^{12}$\,\cmcub\ and gas pressure
higher than 3\,dyn\,cm$^{-2}$ in on-disk flare loops were considered
by \citet{HeinzelKarlicky1987} using the non-LTE modeling of the
\ha\ spectral line. Similar modeling was also performed by
\citet{Svestkaetal1987}. High electron densities of
$10^{12.8}$\,\cmcub\ for the off-limb flare loops were reported in
\citet{Hieietal1983}. They also interpret wide Fe\,{\sc
  xiv}~5303\,\AA\ flare profiles through turbulent velocities of
$30-40$\,\kms. \citet{Hieietal1992} studied a 1989 August 16 X20-class
WL flare and the corresponding off-limb flare loops, concluding that
the WL emission of the apex is due to hydrogen free-bound/free-free
emission. The electron density was estimated to be about
$10^{12-13}$\,\cmcub. Analyzing WL observations of flare loops of a
X2.8-class off-limb flare obtained by SDO/HMI,
\citet{Saint-Hilaireetal2014} estimate the free electron density of
the WL loop system to be as high as $1.8 \times 10^{12}$\,\cmcub.

In contrast, using \ha\ observations of various on-disk and off-limb
flare loops acquired by the {\it Multichannel Subtractive Double-Pass
  spectrograph} (MSDP) at Pic du Midi,
\citet{Heinzeletal1992a,Heinzeletal1992b} and
\citet{Schmiederetal1996b,Schmiederetal1996a} inferred relatively low
electron densities of $2-10 \times 10^{10}$\,\cmcub\ and gas pressures
of the order of $0.1-0.5$\,dyn\,cm$^{-2}$. A 1D radiation-hydrodynamic
model of flare loops by \citet{Tsiklaurietal2004}, based on the
Bastille-Day flare of 2000 July 14, shows that flare loop densities up
to $5 \times 10^{11}$\,\cmcub\ can be achieved \citep[Chapter
  16]{Aschwanden2005}. Recently, \citet{FirstovaPolyakov2017} analyzed
\ha\ spectral measurements of off-limb M7.7-class flare loops, finding
electron densities of $10^{11}$\,\cmcub.

This paper presents an analysis of high-resolution imaging
spectroscopy data of off-limb cool loops of the X8.2-class flare
acquired by the {\it Swedish 1-m Solar Telescope}
\citep[SST;][]{Scharmeretal2003a,Scharmeretal2003b}. We use the 1D
non-LTE radiative-transfer code based on the Multilevel Accelerated
Lambda Iterations technique \citep[MALI, see e.\,g.,][]{Heinzel1995}
and invert the SST spectra using an extensive grid of models. These
inversions are used to obtain diagnostic information on key plasma
parameters of selected bright patches at the flare loop apex at three
times during the gradual phase, focusing mainly on electron densities
and gas pressures.

\section{Observations and data reduction}

\subsection{Target, event, and observational setup}
\label{subs1}

The target of opportunity for SST observations of 2017 September 10
was AR NOAA 12673, which at that time was just behind the western
limb. The target continued with its previous high flare activity
\citep[see e.\,g.,][]{Verma2018,Romanoetal2019} producing at 15:35\,UT
the SOL2017-09-10T16:06 X8.2-class off-limb flare, with the flare peak
at 16:06\,UT \citepalias[Figure\,1]{Kuridzeetal2019}. SST observations
of the event commenced at 16:07:21\,UT and continued until
17:58:37\,UT with the central heliocentric coordinates of the SST
field-of-view (FoV) at ({\it x, y}) = (947\arcsec, $-138$\arcsec)
during the start of observing. The SST observations were carried out
with the CRisp Imaging SpectroPolarimeter
\citep[CRISP;][]{Scharmer2006,Scharmeretal2008} and the CHROMospheric
Imaging Spectrometer \citep[CHROMIS;][]{Lofdahletal2018}, both based
on high-performance dual Fabry-P\'{e}rot interferometers.

The CRISP data comprises imaging spectropolarimetry in 21 line
positions over the \cawav\ line profile, sampled from $-1.75$ to
+1.75\,\AA\ at positions $\pm 1.75$, $\pm 0.945$, $\pm 0.735$, $\pm
0.595$, $\pm 0.455$, $\pm 0.35$, $\pm 0.28$, $\pm 0.21$, $\pm 0.14$,
$\pm 0.07$, and 0.0\,\AA\ from line center (hereafter, unless
specified otherwise, when referring to the \ca\ line we mean the
\cawav\ line). Each spectral scan of \ca\ had an acquisition time of
16\,s but the time series cadence is 33\,s due to the inclusion of
scans in the Fe\,{\sc i}~6302\,\AA\ photospheric line. The spatial
sampling is 0\farcs057\,pixel$^{-1}$ over a square FoV of $41 \times
41$\,Mm$^2$. We analyse the \ca\ Stokes {\it I} profiles here - the
Stokes {\it Q,U,V} profiles are presented in
\citetalias{Kuridzeetal2019}. The full width at half maximum (FWHM) of
the CRISP transmission profile at the \ca\ line is 107.3\,m\AA\ with a
prefilter FWHM of 9.3\,\AA\ \citep{delaCruzRodriguezetal2015}. The
CRISP data are processed by the CRISPRED reduction pipeline
\citep{delaCruzRodriguezetal2015} and reconstructed with the
Multi-Object Multi-Frame Blind Deconvolution
\citep[MOMFBD;][]{Lofdahl2002,vanNoortetal2005}.

Simultaneous observations were taken with the CHROMIS imaging
spectrometer, which observes the blue part of the spectrum in the
range $3900 - 4900$\,\AA. The CHROMIS observations comprise spectral
imaging in the hydrogen \hb~4861\,\AA\ and Ca\,{\sc ii}\,H~3968.5 \&
K~3933.7\,\AA\ lines plus one position in the continuum at
4000\,\AA. CHROMIS also contains a wide-band imaging system, having an
identical beam as the Fabry-P\'{e}rot interferometers, with a filter
centered at 4845.5\,\AA\ with a FWHM of 6.5\,\AA\ (see Figures~2 and 3
and Table~1 in \citet{Lofdahletal2018}). Flare-loop emission in this
wide band will be studied in a future paper. Temporal cadence of the
time series is 20\,s and the spatial sampling of
0\farcs0375\,pixel$^{-1}$ over a rectangular FoV of about $45 \times
30$\,Mm$^2$. The \hb\ line profile was sampled from $-1.2$ to
+1.2\,\AA\ at positions $\pm 1.2$, $\pm 1.0$, $\pm 0.8$, $\pm 0.7$,
$\pm 0.6$, $\pm 0.5$, $\pm 0.4$, $\pm 0.3$, $\pm 0.2$, $\pm 0.1$, and
0.0\,\AA\ from line center. The CHROMIS transmission FWHM at \hb\ is
$\approx 130$\,m\AA\ with a prefilter FWHM of
4.8\,\AA\ \citep{Lofdahletal2018}. CHROMIS data were processed using
the CHROMISRED reduction pipeline, which includes MOMFBD image
restoration and absolute intensity calibration
\citep{Lofdahletal2018}.

\begin{figure}
\includegraphics[width=\textwidth]{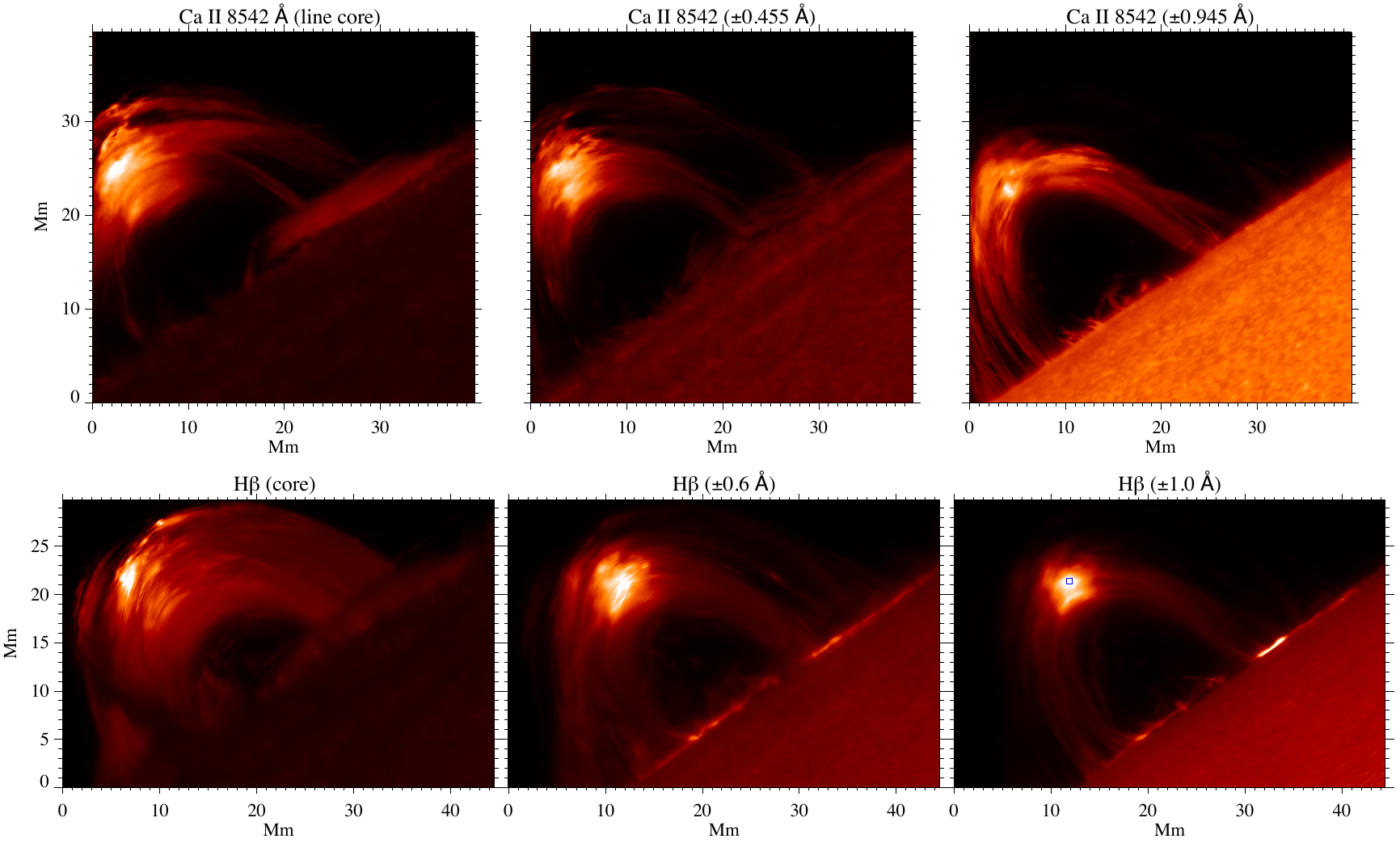}
\caption{Overview of SST observations of the flare loops on 2017
  September 10 at the western limb. Each image is byte-scaled
  independently. Top: CRISP \cawav\ images at 16:28:27\,UT, line core
  image (left), composites of near-wing images at $\Delta\lambda = \pm
  0.455$\,\AA\ (middle) and wing images at $\Delta\lambda = \pm
  0.945$\,\AA\ (right). Bottom: CHROMIS \hb\ images at 16:26:43\,UT,
  line core image (left), composites of near-wing images at
  $\Delta\lambda = \pm 0.6$\,\AA\ (middle) and wing images at
  $\Delta\lambda = \pm 0.1$\,\AA\ (right). Bottom right frame: the
  blue box marks the bright patch over which were averaged profiles
  shown in the middle column of Figure~\ref{fig4}.}
\label{fig1}
\end{figure}

\subsection{Spatial alignment of data}

\begin{figure}
\centering
\includegraphics[width=0.9\textwidth]{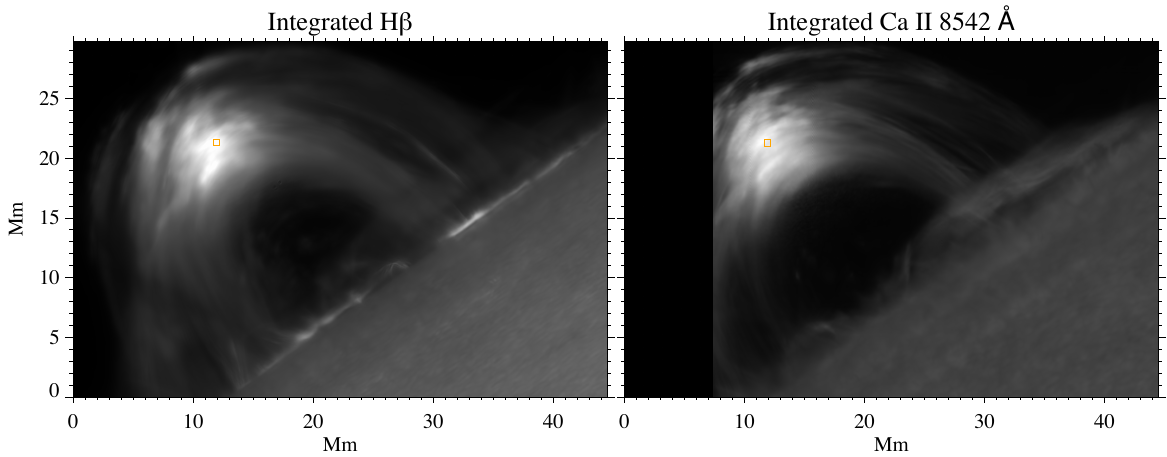}
\caption{Line-integrated emissions of flare loops. The orange boxes
  are co-spatial with the blue box in Figure~\ref{fig1} (bottom right
  frame). They mark the bright patch over which were averaged profiles
  shown in the middle column of Figure~\ref{fig4}.}
\label{fig2}
\end{figure}

Figure~\ref{fig1} shows an overview of the MOMFBD-processed SST
spectral imagery of the flare loops acquired on 2017 September 10. The
CRISP (top frames) and CHROMIS (bottom) images were acquired at
16:28:27\,UT and 16:26:43\,UT respectively. Although made close in
time, their FoV centers are not spatially co-aligned. Due to highly
variable and less-than-optimum seeing, only three quasi-simultaneous
pairs of line profile scans, taken at the moment of the best viewing
conditions at $\approx$\,(16:25, 16:27, 16:28)\,UT, are used for
analysis. To co-align the loop multi-spectral imagery, images in
wavelength-integrated intensities over \hb\ and \ca\ profiles are
created for the three times of observation, allowing clearly
recognizable common features to be identified, mainly at the bright
loop apex (Figure~\ref{fig2}). Through cross-correlation, a
satisfactory spatial alignment of \ca\ and \hb\ data is achieved as
shown in Figure~\ref{fig2}.

\newpage
\subsection{Data radiometric calibration}

\begin{figure}
\centering
\includegraphics[width=0.49\textwidth]{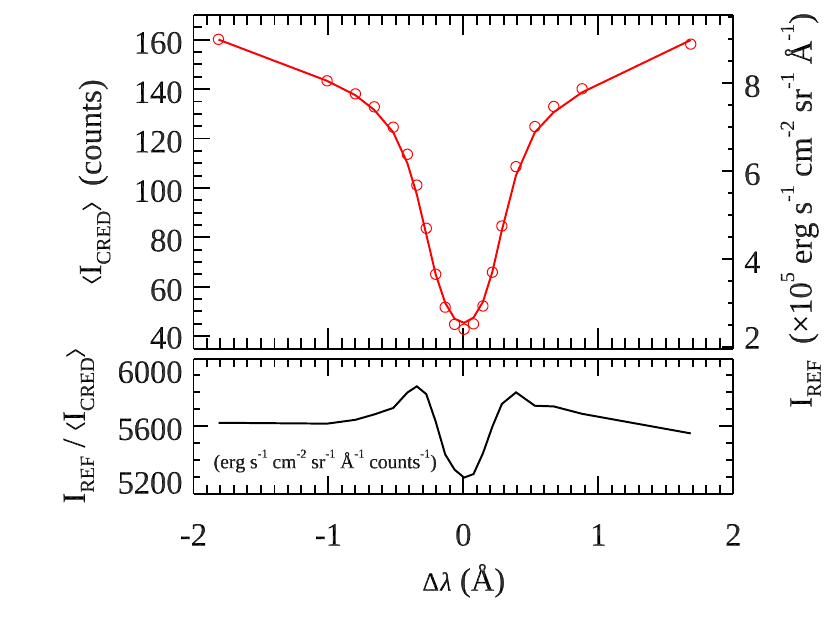}
\includegraphics[width=0.49\textwidth,bb=0 -10 240 161]{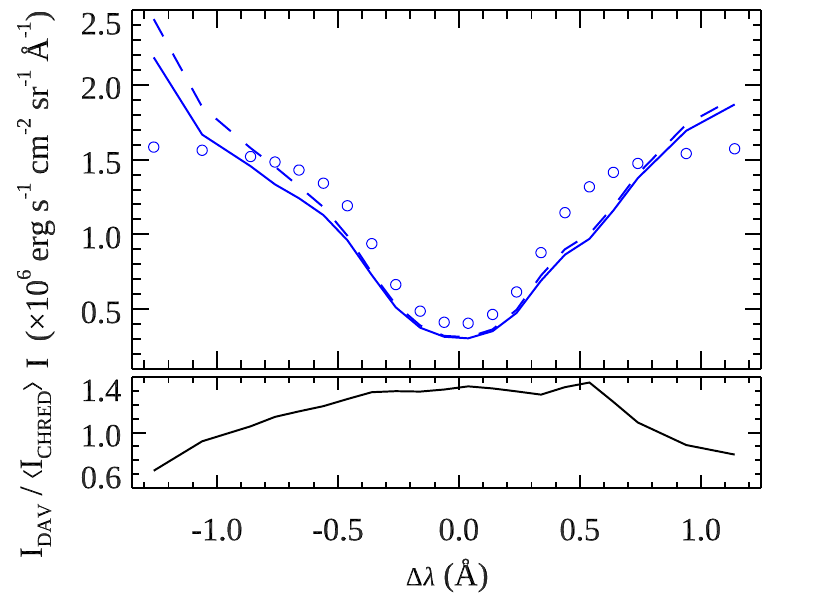}
\caption{Radiometric calibration profiles. Left: \cawav\ profile
  yielded by CRISPRED $\langle I_{\rm CRED} \rangle$ (solid red line)
  averaged over a quiet-Sun area at the directional cosine $\mu
  \approx 0.20$. The reference profile $I_{\rm REF}$ at the same $\mu$
  extrapolated from \citet{Linskyetal1970} and convolved with CRISP
  transmission (red circles). The bottom frame shows their ratio used
  in calibrating CRISP data. Right: \hb\ profile yielded by CHROMISRED
  averaged over a quiet-Sun area at $\mu \approx 0.18$ over three
  given moments $\langle I_{\rm CHRED} \rangle$ (solid blue line) and
  for a single moment at 16:26:43\,UT (dashed). The reference profile
  $I_{\rm DAV}$ at the same $\mu$ taken from \citet{David1961}
  convolved with CHROMIS transmission (blue circles). The bottom frame
  shows their ratio used in re-calibrating CHROMIS data.}
\label{fig3}
\end{figure}

\begin{figure}
\includegraphics[width=\textwidth]{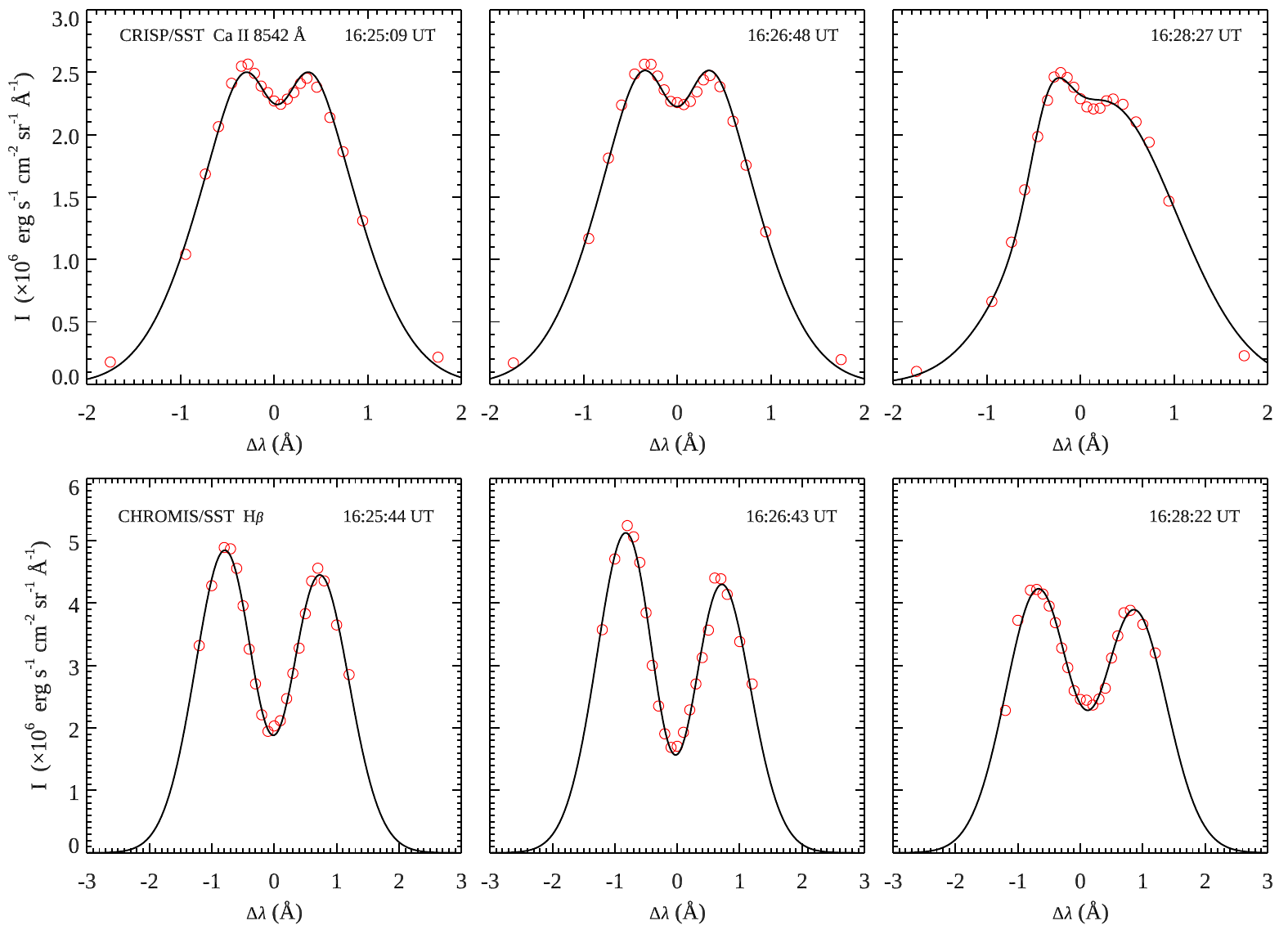}
\caption{Observed line profile intensities (red circles) and their
  double-Gaussian fits (black lines) averaged over selected bright
  patches (boxes in Figures~\ref{fig1} and \ref{fig2}). Profiles in
  the same column are co-spatial but columns do not trace the same
  patch.}
\label{fig4}
\end{figure}

Diagnostics of plasma parameters through the non-LTE
radiative-transfer computations requires careful calibration of
spectral data in absolute intensity units.
Observed data, yielded by the CRISPRED pipeline, are represented by a
\ca\ profile $\langle I_{\rm CRED} \rangle$ (Figure~\ref{fig3}: top
left frame) taken as the spatial average over the $165 \times
145$\,px$^2$ = $9\farcs4 \times 8\farcs3 \approx 6.8 \times
6.0$\,Mm$^2$ rectangle centered at a quiet-Sun area
(Figure\,\ref{fig1}: top frames) at ({\it x, y}) = (34, 4)\,Mm at 
the direction cosine $\mu
= 0.2032$. The average profile $\langle I_{\rm CRED} \rangle$ is
compared with the reference profile $I_{\rm REF}$ taken from
\citet{Linskyetal1970}. The latter profile is: (i) extrapolated for
the given $\mu$, (ii) intensity calibrated by the
disk-center absolute continuum intensity taken from \citet{Cox2000},
(iii) corrected for limb darkening by the formula given therein, and
finally (iv) convolved with the transmission profile of the the CRISP
Fabry-P\'{e}rot etalons provided by J.~de la Cruz Rodr\'{\i}guez
(2017, private communication). The ratio of the reference profile and
the observed average renders a calibration profile $I_{\rm REF} /
\langle I_{\rm CRED} \rangle$ (Figure~\ref{fig3}: bottom left frame)
allowing the conversion of the \ca\ flare loop profiles from digital
to absolute units (Figure\,\ref{fig4}: top frames).

The CHROMISRED reduction pipeline implicitly performs calibration in
disk-center absolute intensity units using the Hamburg disk center
spectral atlas \citep{Neckel1999}, disregarding position angle of
CHROMIS science data \citep[Subsection~4.3.]{Lofdahletal2018}. We verify
the calibration by the \hb\ profile taken as a spatial average over
the $400 \times 200$\,px$^2$ = $15\farcs0 \times 7\farcs5 \approx 10.9
\times 5.4$\,Mm$^2$ rectangle centered at a quiet-Sun area
(Figure\,\ref{fig1}: bottom frames) at ({\it x, y}) = (37, 3.5)\,Mm at
the direction cosine $\mu = 0.1789$.  The \hb\ disk intensity is
corrected for limb darkening by the formula of \citet{Cox2000}, and
further corrected for air mass increase through a disc-center flat
field measurement at 14:45\,UT and science measurements at
$\approx$\,16:27\,UT. The CHROMISRED pipeline used the flat fields in
radiometric calibration. The resulting \hb\ profile $\langle I_{\rm
  CHRED}(\lambda,\mu) \rangle$ is compared with the \hb\ profile
$I_{\rm DAV}(\lambda,\mu)$ taken from \citet{David1961}
(Figure~\ref{fig3}: top right frame). The latter profile is
interpolated for the given $\mu$ and further
processed with identical steps to CRISP (ii)-(iv), but with a
transmission profile provided by (M.~L\"{o}fdahl 2019, private
communication). The coefficients of limb darkening of 0.399, the air
mass change of 1.083, and the ratio $I_{\rm
  DAV}(\lambda,\mu)$/$\langle I_{\rm CHRED}(\lambda,\mu) \rangle$
(Figure~\ref{fig3}: bottom right frame) are applied in re-calibrating
the \hb\ flare loop profiles yielded by CHROMISRED
(Figure\,\ref{fig4}: bottom frames).  

\subsection{Flare loop appearance}

\begin{figure}
\centering
\includegraphics[width=0.7\textwidth]{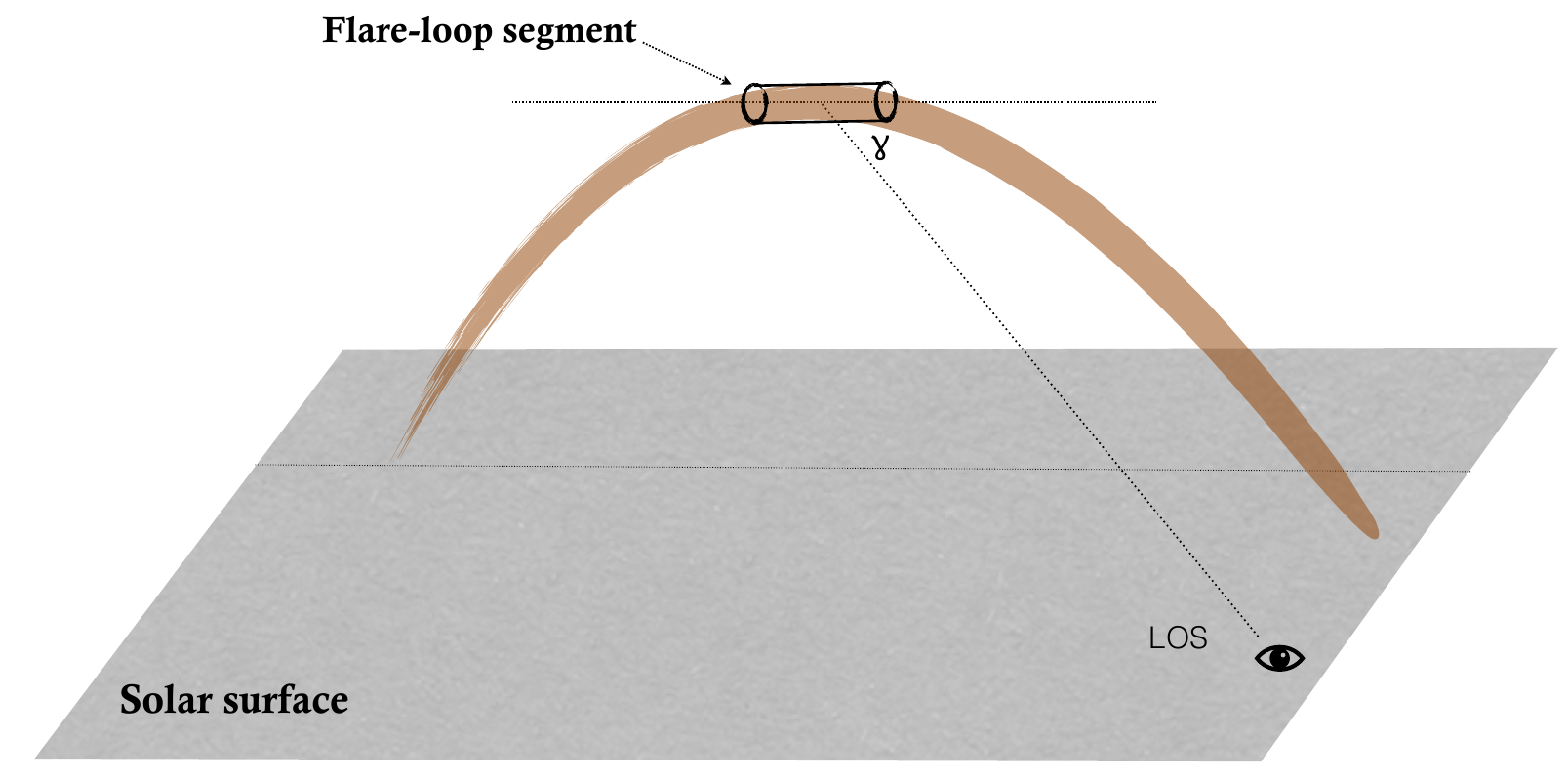}
\caption{Schematic rendering of the flare loop showing the basic
  off-limb geometry, the line of sight LOS, and the viewing angle
  $\gamma$. Quasi-cylindrical flare-loop segment represents the bright
  patch in Figures~\ref{fig1} and \ref{fig2}. In our non-LTE modeling,
  we replace the flare-loop segment by a 1D vertical slab, which is
  isothermal and isobaric. Its width corresponds to the width of this
  cylindrical segment.}
\label{fig5}
\end{figure}

The line core images show very prominent bright apices of axially
quasi-symmetric flare loops (Figure~\ref{fig1}: left column), whose
axes lie almost along the line of sight (LoS), with the right leg and
footpoint displaced towards the observer (Figure~\ref{fig5};
\citetalias[Figure~12]{Kuridzeetal2019}). The loop legs appear faint
due to significant Doppler shifts
\citepalias[Figure~11a]{Kuridzeetal2019} and presumably due to
  lower density compared to the apex. The right leg is partially
obscured behind a prominence that is situated at the limb.

The \hb\ near-wing and wing images show bright narrow rims at the loop
footpoints (Figure~\ref{fig1}: middle and right frames at the bottom),
which closely follow the limb. We interpret them as \hb\ flare ribbons
seen edge-on. They are absent in the \hb\ and \ca\ core images due to
the opaque chromospheric canopy with the maximum opacity at the line
cores (Figure~\ref{fig1}: left column). Remarkably, the \hb\ ribbon at
the right leg is substantially brighter than the ribbon at the left
leg (Figure~\ref{fig1}: bottom right frame). It suggests that the
right leg and ribbon are closer to the observer than the left one
(Figure~\ref{fig5}; \citetalias[Figure~12]{Kuridzeetal2019}), which is
likely partially obscured behind the limb. The blueshifts and
redshifts detected in the right and left legs, respectively, support
this scenario \citepalias[Figure~11a]{Kuridzeetal2019}. However, the
ribbons are absent in the \ca\ near-wing and wing images
(Figure~\ref{fig1}: middle and right frames at the top). This behavior
of footpoints will be analyzed in another paper.

\subsection{Basic characteristics of selected line profiles}

Figure~\ref{fig4} shows the observed line profiles that are selected
for inversion through 1D non-LTE modeling. They are averaged over
$\approx 0\farcs7 \times 0\farcs7$-wide boxes; rendered in
Figures~\ref{fig1} and \ref{fig2} in blue and orange, respectively;
showing the strongest emission in Figure~\ref{fig2}. The red circles
in Figure~\ref{fig4} illustrate the wavelength sampling of the
observed profiles given in Subsection~\ref{subs1}. Thus, the profiles
data represent the brightest and very likely the densest region of the
loop apex. This choice guarantees the best measurement signal-to-noise
ratio. The selected bright patches are situated around 17\,Mm above
the solar photosphere. At this altitude, an angle between LoS and the
local normal to the loop plane is estimated to be $\approx
25$\degr\ \citepalias[Figure~11b]{Kuridzeetal2019}. The columns of
Figure~\ref{fig4} show strictly co-spatial line profiles. The profiles
in subsequent columns correspond to the selected bright regions
(boxes) at the three different times, but should not be interpreted in
terms of a time evolution of the same loop plasma.  A striking feature
of the \hb\ profiles (Figure~\ref{fig4}: bottom frames) is a deep
central reversal surrounded by slightly asymmetric peaks suggesting a
Doppler shift gradient along LoS
\citep{Kuridzeetal2015,Kuridzeetal2016,Kuridzeetal2017}. The reversals
and signatures of peak asymmetries are also apparent in the
\ca\ profiles (Figure~\ref{fig4}: top frames).

\subsection{Double-Gaussian fitting and symmetrization}

Because our observations were not tailored for very broad profiles
occurring in solar flares (Subsection~\ref{subs1}) and our non-LTE
modeling assumes a static plasma without any bulk motions, thus
yielding symmetric profiles, the observed profile intensities (the red
circles in Figure~\ref{fig4}) were further processed to be comparable
with the results of modeling. The observed intensities of both lines
$I(\lambda)$ were fitted by the double-Gaussian model function. The
double-Gaussian fits are displayed in Figure~\ref{fig4} by the black
solid lines. For the fitting the SolarSoft function
\texttt{mpfitfun.pro} is used. The function calls the core procedure
\texttt{mpfit.pro}, which performs Levenberg-Marquardt least-square
minimization of a $\chi^2$ merit function
\citep{Markwardt2009,More1978,MoreWright1993}. Further, the fits are
symmetrized with respect to the zero wavelength by taking their
average $0.5 [ I(\Delta\lambda)+I(-\Delta\lambda) ]$. This compensates
for the peak intensity differences and possible small profile
asymmetries and shifts with respect to the rest line-center
wavelength.  The symmetric fits within the scanning ranges of $\pm
1.75$\,\AA\ and $\pm 1.2$\,\AA\ of CRISP and CHROMIS, respectively,
are shown in Figures~\ref{fig6} and \ref{fig7} by the black solid
lines.

As explained, the symmetric fits are used as a proxy of the real
profiles in the inversion. It is important to consider the effect of
symmetrization on their basic characteristics and their
spatio-temporal variability.
Comparing the integrated intensities $E$, the reversal depths $r$ of
the observed \ca\ profiles, and their symmetric fits shows differences
typically less than $\approx 6$\%. This confirms the validity of using
symmetric fits as proxies of the observed profiles in the
inversion. Despite using just three $E$ and $r$ values of the
\ca\ profiles, their spatio-temporal variability over bright patches
at loop apex is less than $\approx 9$\%.
On the other hand, the \hb\ profiles and their symmetric fits manifest
larger variability. Obviously, due to missing wings in the observed
\hb\ profiles, their integrated intensities are smaller than the
values inferred by the symmetric fits. But as proven earlier by the
\ca\ profiles, the symmetric \hb\ fits can be considered as a good
proxy in the inversion. This is also confirmed by the close similarity
of the observed \hb\ reversal depths $r$ and the depths of the
symmetric fits. The symmetric fits indicate a decrease of \hb\ and
\ca\ integrated intensities over the loop apex within the time
interval of $\approx 3$\,min, but this is a too small statistical
sample, and too short time interval for any firm conclusions.

\section{Description of the inversion process}

The inversion is performed by exploring a multidimensional space of
plasma parameters using several grids of models. Each model is
characterized by a unique set of four free parameters (see
Subsection~\ref{mali}). A grid of models comprises an extensive set of
individual models, usually more than several tens of thousands, whose
parameters change by regular increments. The grids differ by widths of
parameter ranges, sizes of sampling steps, and number of models. For
each model grid a library of synthetic line profiles is
computed. These are convolved with the instrumental characteristics in
order to simulate the observed profiles.  Comparing the library of
synthetic profiles $I_{\rm syn}(\lambda_i)$ with the observed profiles
$I_{\rm obs}(\lambda_i)$ identifies a model that yields the best
match. A similarity of the profiles, sampled in the $N$ wavelengths
$\lambda_i~(i = 1 \dots N)$ with the measurement uncertainties of
$\sigma(\lambda_i)$, is quantified by the $\chi^2$ merit function
weighted by $\sigma(\lambda_i)$. The function is computed within the
wavelength range sampled by the CHROMIS \hb\ observations but only up
to 2.25\,\AA\ for \ca. Its global minima define the optimal model
parameters that represent a plausible solution of the problem.  In the
following, we assume, for both the \ca\ and \hb\ lines, constant and
wavelength-independent value of $\sigma = 2 \times
10^{-6}$\,erg\,s$^{-1}$\,cm$^{-2}$\,sr$^{-1}$\,Hz$^{-1}$, which
converts into $0.8 \times 10^5$ and $2.5 \times
10^5$\,erg\,s$^{-1}$\,cm$^{-2}$\,sr$^{-1}$\,\AA$^{-1}$ for \ca\ and
\hb, respectively. Double- and single-line inversions are performed
separately. For the former, the merit function is evaluated
simultaneously for both lines concatenating their $I_{\rm
  obs}(\lambda_i)$ into one vector as well as $I_{\rm
  syn}(\lambda_i)$.  For the latter the merit function is evaluated
separately for each line. The corresponding results are distinguished
by the symbols \ddag\ and \dag\ in Tables~\ref{tab1}, \ref{tab3}, and
\ref{tab4}. Both, $I_{\rm obs}(\lambda_i)$ and $I_{\rm
  syn}(\lambda_i)$ are calibrated in units of
erg\,s$^{-1}$\,cm$^{-2}$\,sr$^{-1}$\,Hz$^{-1}$ and resampled on a
denser wavelength grid with the step of 0.01\,\AA\ yielding $N$(\ca) =
226 and $N$(\hb) = 121 intensity samples for the $\chi2$ computation.

Table~\ref{tab1} compares the integrated intensities $E$ and the
central reversal depths $r$ of the observed profiles, their
double-Gaussian and symmetric fits, and the synthetic profiles. For
the latter, the line center optical thicknesses $\tau_0$ are also
listed. The reversal depth $r$ is defined as: $r = {\rm mean}(I_{\rm
  b},I_{\rm r})/I_{\rm min}$, where $I_{\rm b}$ and $I_{\rm r}$ are
the intensities of the blue and red peaks, respectively, around the
reversal minimum intensity $I_{\rm min}$. 

  \begin{table}
    \centering
    \caption{Wavelength-integrated intensities $E$ in the units
      $10^6$\,erg\,s$^{-1}$\,cm$^{-2}$\,sr$^{-1}$, the central
      reversal depths $r$, and the line center optical thickness
      $\tau_0$ of the \cawav\ and \hb\ lines.}
    \label{tab1}
      \begin{tabular}{cc cc Dc Dc c@{\hspace{-2pt}} D@{\hspace{-2pt}}D c r@{\hspace{2pt}}r}
      \hline
      \hline
      \decimals
        \multirow{2}{*}{Spectral line} & \multirow{2}{*}{hh:mm:ss\,UT} & \multicolumn2c{Observations} & \multicolumn3c{Double-Gaussian fit} & \multicolumn3c{Symmetric fit} & & \multicolumn7c{Synthetic profile\tablenotemark{a}} \\[-1.4mm]
                                       &                               &       $E$    & $r$           &        \multicolumn2c{\phn{}$E$} &  $r$   &         \multicolumn2c{\phn{}$E$} & $r$         & & \multicolumn4c{$E$} & $r$ & \multicolumn2c{$\tau_0$} \\
      \hline
      \multirow{6}{*}{\cawav} & \multirow{2}{*}{16:25:09} & \multirow{2}{*}{4.84} & \multirow{2}{*}{1.12} & \multirow{2}{*}{4.90} & \multirow{2}{*}{1.12} & \multirow{2}{*}{4.90} & \multirow{2}{*}{1.12} & \ddag & 4.65 & (4.77) & 1.01 (1.00) &  4.5 & (3.7) \\ 
                              &                           &                       &                       &          .             &         &       .     &              &  \dag & 4.80 & (4.85) &  1.00 (1.00)  &  3.0 & (2.3) \\ 
      \cline{11-18} 
                              & \multirow{2}{*}{16:26:48} &  \multirow{2}{*}{4.92}  &  \multirow{2}{*}{1.12} & \multirow{2}{*}{4.92} & \multirow{2}{*}{1.13} & \multirow{2}{*}{4.92}   &  \multirow{2}{*}{1.13}   &  \ddag & 4.69 & (4.75) &  1.01  (1.01) &  4.7 & (4.6) \\ 
                             &                            &                        &                       &          .            &         &       .      &             & \dag & 4.84 & (4.87) &  1.00 (1.00) &  3.1 & (2.3) \\ 
      \cline{11-18} 
                           & \multirow{2}{*}{16:28:27} &  \multirow{2}{*}{4.49}  &  \multirow{2}{*}{1.08}  &  \multirow{2}{*}{4.69}  &  \multirow{2}{*}{1.03}  &  \multirow{2}{*}{4.69}   &  \multirow{2}{*}{1.02}    &  \ddag & 4.34 & (4.37) &  1.00 (1.00) & 3.8 & (3.6) \\ 
                             &                             &                       &                       &          .           &          &          .    &                   & \dag & 4.48 & (4.48) &  1.00  (1.00) & 2.3 & (1.7) \\  
      \hline
      \multirow{6}{*}{\hb} & \multirow{2}{*}{16:25:44} & \multirow{2}{*}{8.51}  & \multirow{2}{*}{2.43}  &  \multirow{2}{*}{10.66}  &  \multirow{2}{*}{2.46}  &  \multirow{2}{*}{10.66}   & \multirow{2}{*}{2.46}    & \ddag & 12.79 & (12.18) & 1.92 (1.78) & 69.5 & (43.7) \\ 
                           &                               &                        &                       &          .         &            &          .    &                    & \dag & 13.12 & (12.50) & 2.06 (1.92) & 117.0 & (69.6) \\  
      \cline{11-18} 
                           & \multirow{2}{*}{16:26:43} &  \multirow{2}{*}{8.37}  &  \multirow{2}{*}{2.86}  &  \multirow{2}{*}{10.52}   &  \multirow{2}{*}{2.95}  &  \multirow{2}{*}{10.52}   &  \multirow{2}{*}{2.95}    & \ddag & 12.85 & (12.13) & 1.95 (1.85) & 74.7 & (56.0) \\  
                           &                               &                        &                       &          .         &             &         .     &                     & \dag & 13.25 & (12.50) & 2.05 (1.92) &  109.0 & (69.6) \\ 
      \cline{11-18}  
                           & \multirow{2}{*}{16:28:22} &  \multirow{2}{*}{8.04}  &  \multirow{2}{*}{1.71}  &  \multirow{2}{*}{10.30}   &  \multirow{2}{*}{1.77}  &  \multirow{2}{*}{10.30}   &  \multirow{2}{*}{1.77}    & \ddag & 12.00 & (10.49) & 1.92 (1.78) &  59.4 & (37.1) \\   
                           &                               &                        &                        &         .         &            &          .      &                    & \dag & 11.83 & (10.30) & 2.06 (1.74) &  99.4 & (28.5) \\  
      \hline
    \end{tabular}
\tablenotetext{a}{ \ddag\ -- double-line inversion, \dag\ -- single-line inversion. Unparenthesized and parenthesized data correspond to the parameters of model~1 and 2 in Table~\ref{tab3}, respectively.} 
  \end{table}

\subsection{MALI radiative-transfer code}
\label{mali}

For synthetic profile calculations of \ca\ and \hb\ the MALI technique
of \citet{RybickiHummer1991} is used, as it was implemented for
prominence-like structures by \citet{Heinzel1995}.  The static version
solves the radiative transfer in an isothermal and isobaric vertical
slab (of geometrical thickness $D$) without any bulk motions. A bright
patch at the flare loop apex is represented by such 1D slab located
above the solar surface at a pre-defined height $h$ and irradiated
symmetrically from both sides by an ambient solar-disk radiation
(Figure~\ref{fig5}). The geometry of the problem is given by the angle
$\theta$ between LoS and normal to the slab $\theta = 90\degr -
\gamma$. For the region of altitude $h \approx 17$\,Mm, the viewing
angle $\gamma$ is estimated to be $\approx 65\degr$
\citepalias[Figure~11b]{Kuridzeetal2019}, thus $\theta \approx
25\degr$. Free input model parameters for the synthetic profile
calculations are temperature $T$, gas pressure $\pg$, slab geometrical
thickness $D$, and microturbulent velocity $\vt$ (see
Table~\ref{tab2}). When convergence of the MALI iterations is
achieved, the code yields, among others, the synthetic intensities of
line profiles in erg\,s$^{-1}$\,cm$^{-2}$\,sr$^{-1}$\,Hz$^{-1}$, their
integrated intensities $E$, line-center optical thickness $\tau_0$,
and the electron densities in the center $\nec$ and at the surface
$\nes$ of the slab.  The synthetic \ca\ profiles are computed in 37
wavelength samples from 0.0 to 2.25\,\AA\ with a variable wavelength
step ranging from 0.025\,\AA\ in the core to 0.25\,\AA\ in the far
wings. The \hb\ profiles are computed in 42 points from 0 to
3.61\,\AA\ also with a variable step ranging from 0.045\,\AA\ in the
core to 0.36\,\AA\ in the far wing.

We initially solve the non-LTE problem for hydrogen, in order to get
the electron densities and compute the radiation field in Lyman lines
and the continuum. The latter are then used for evaluation of the
\ca\ photoionization rates \citep[see][]{GouttebrozeHeinzel2002}. The
non-LTE MALI computation of \ca\ follows. Five-level plus continuum
hydrogen and \ca\ atom models are considered (the population of
Ca~{\sc i} is negligible). Depending on the formation properties of
spectral lines, complete or partial redistribution is used.  Because
the Ca abundance in flare loops is not well understood, the
photospheric Ca abundance of $2.19\times10^{-6}$ ($\log\varepsilon =
6.34$) and the coronal Ca abundance $4.36\times10^{-6}$
($\log\varepsilon = 6.64$) recommended in \citet{Grevesseetal2010} and
\citet{Schmeltzetal2012}, respectively, are adopted.

  \begin{table}
    \centering
    \caption{Parameters of model grids employed in the inversion process.}
    \label{tab2}
    \begin{tabular}{l r@{\,--\,}lcc r@{\,--\,}lcc r@{\,--\,}lcc}
      \hline
      \hline
      \multirow{2}{*}{Parameter} & \multicolumn4c{Extended coarse grid} & \multicolumn4c{Fine grid 1} & \multicolumn4c{Fine grid 2} \\
                & \multicolumn2c{Range} & Step\,size & No.\,of\,points & \multicolumn2c{Range} & Step\,size & No.\,of\,points & \multicolumn2c{Range} & Step\,size & No.\,of\,points \\
      \hline
      $T$ (kK)                      &  6&20 & 0.5            &    29 &    8&10   & 0.1       & 21 &  8&10   & 0.1 &   21 \\
      $\pg$ (dyn\,cm$^{-2}$)         &  1&20 &  1\phn\phd     &    20 &    7&11   & 0.1       & 41 &  10&15  & 0.1 &   51 \\
      $D$ (Mm)                      &  1&5  &  1\phn\phd     & \phn5 &  4.5&6.0  & 0.1       & 16 &   2&3   & 0.1 &   11 \\
      $\vt$ (\kms)                  & 10&40 & 10\phn\phn\phd & \phn4 &   20&32   & 2\phn\phd & \phn7 &  20&36  & 2\phn\phd & \phn9 \\
      $h$ (Mm), $\theta$ (\degr)    & \multicolumn2c{}  & 10, 0   &    &   \multicolumn2c{}    & 17, 25 &   & \multicolumn2c{}  & 17, 25 & \\
      \hline
      No.\,of\,models               & \multicolumn2c{}  & 11\,600 &    &   \multicolumn2c{}    &  96\,432 &  & \multicolumn2c{}  & 106\,029 & \\      
      \hline
    \end{tabular}
  \end{table}

\subsection{Model grids and inversion process}

\begin{figure}
\includegraphics[width=\textwidth]{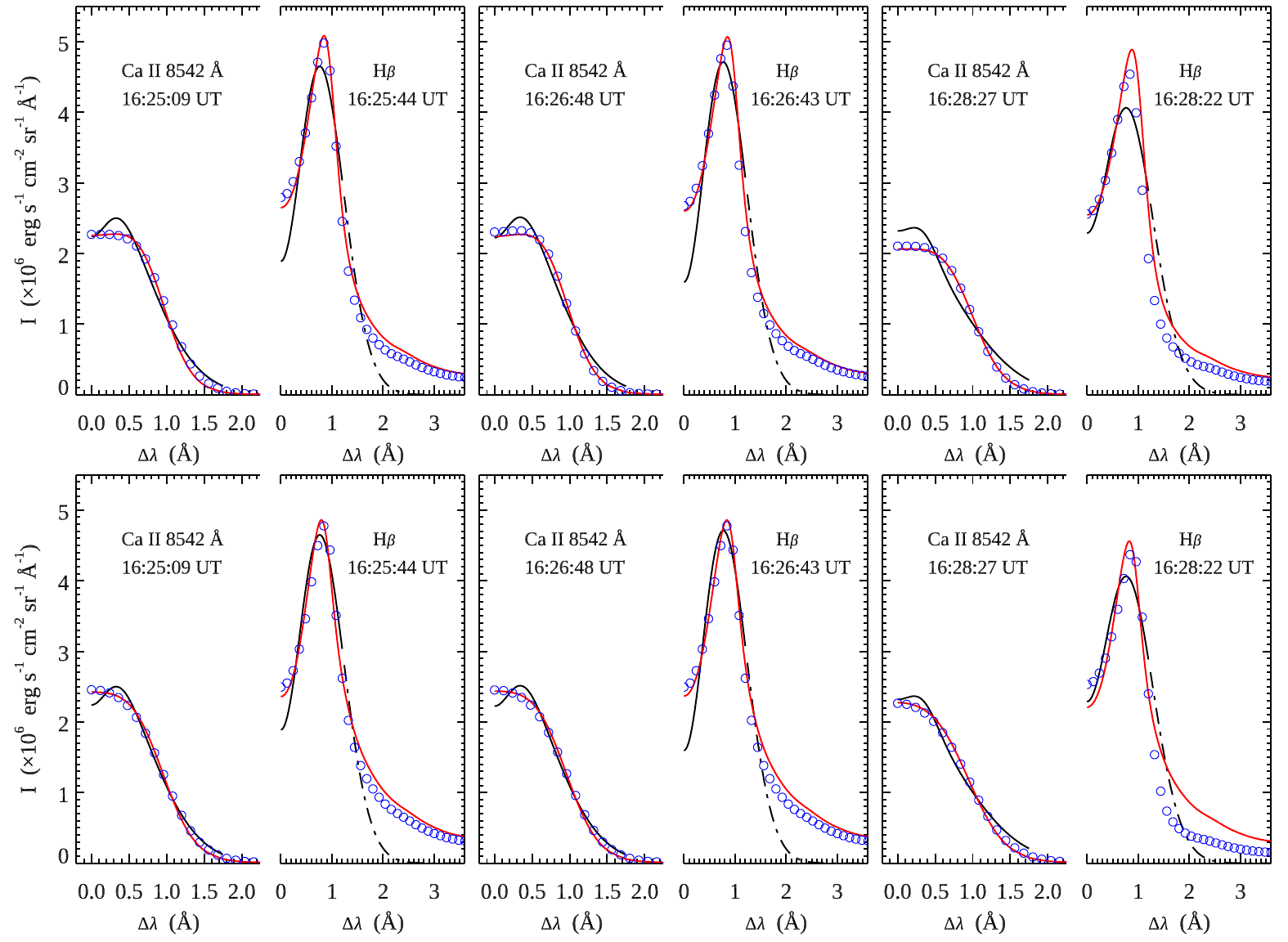}
\caption{Observed (black) and synthetic line profiles corresponding to
  model~1 (red lines) and 2 (blue circles) in Table~\ref{tab3}. The
  observed profiles are represented by their symmetric double-Gaussian
  fits (Figure~\ref{fig4}). The black solid and dash-dotted lines of the
  \hb\ profiles distinguish the wavelength ranges observed by CHROMIS
  and adopted for the Gaussian extrapolation, respectively. Top:
  double-line inversion evaluating the $\chi^2$ merit function for
  concatenated \ca\ and \hb\ profiles. Bottom: single-line inversion
  carried out separately for \ca\ and \hb.}
\label{fig6}
\end{figure}

An initial step in implementing a grid-based inversion is construction
of appropriate model grids. The inversion itself is performed in
several steps by exploring the four-dimensional parameter space $(T,
\pg, D, \vt)$ using the grids of models, which differ by extension of
parameter ranges and sizes of sampling steps. In this approach, the
merit function becomes a multivariate function $\chi^2(T, \pg, D,
\vt)$ of four parameters, which are assumed to be
independent. Firstly, we construct an extended but coarse model grid
(Table~\ref{tab2}) providing a rough estimate of the solution for the
photospheric Ca abundance. Once the minimum of the merit function is
found at the coarse resolution, the search is extended by narrowing
the ranges and refining the grid's mesh around the best-match location
of the first trial. The fine grids, employed in the next trials, are
shown in Table~\ref{tab2}. The best fits for the photospheric Ca
abundance are shown in Figure~\ref{fig6} and the corresponding line
profile characteristics $E$, $r$, $\tau_0$ and resulting model
parameters $T$, $\pg$, $D$, $\vt$, $\nes$, $\nec$ are shown,
respectively, in Tables~\ref{tab1} and \ref{tab3}, where the
parenthesized values correspond to the fine grid~2. An effect of the
Ca abundance increase in cool flare loops is examined by repeating the
inversion using the fine grids in Table~\ref{tab2} with the synthetic
\ca\ profiles computed for the coronal Ca abundance. The resulting
best-fit parameters are given in Table~\ref{tab4} omitting the results
of the \hb\ single-line inversion (which are identical with the values
\dag\,\hb\ in Table~\ref{tab3}). The quality of the resulting fits is
quantified by the $\chi^2$ values given in the last column of
Tables~\ref{tab3} and \ref{tab4} in $1\sigma$ units.

\begin{figure}
\includegraphics[width=\textwidth]{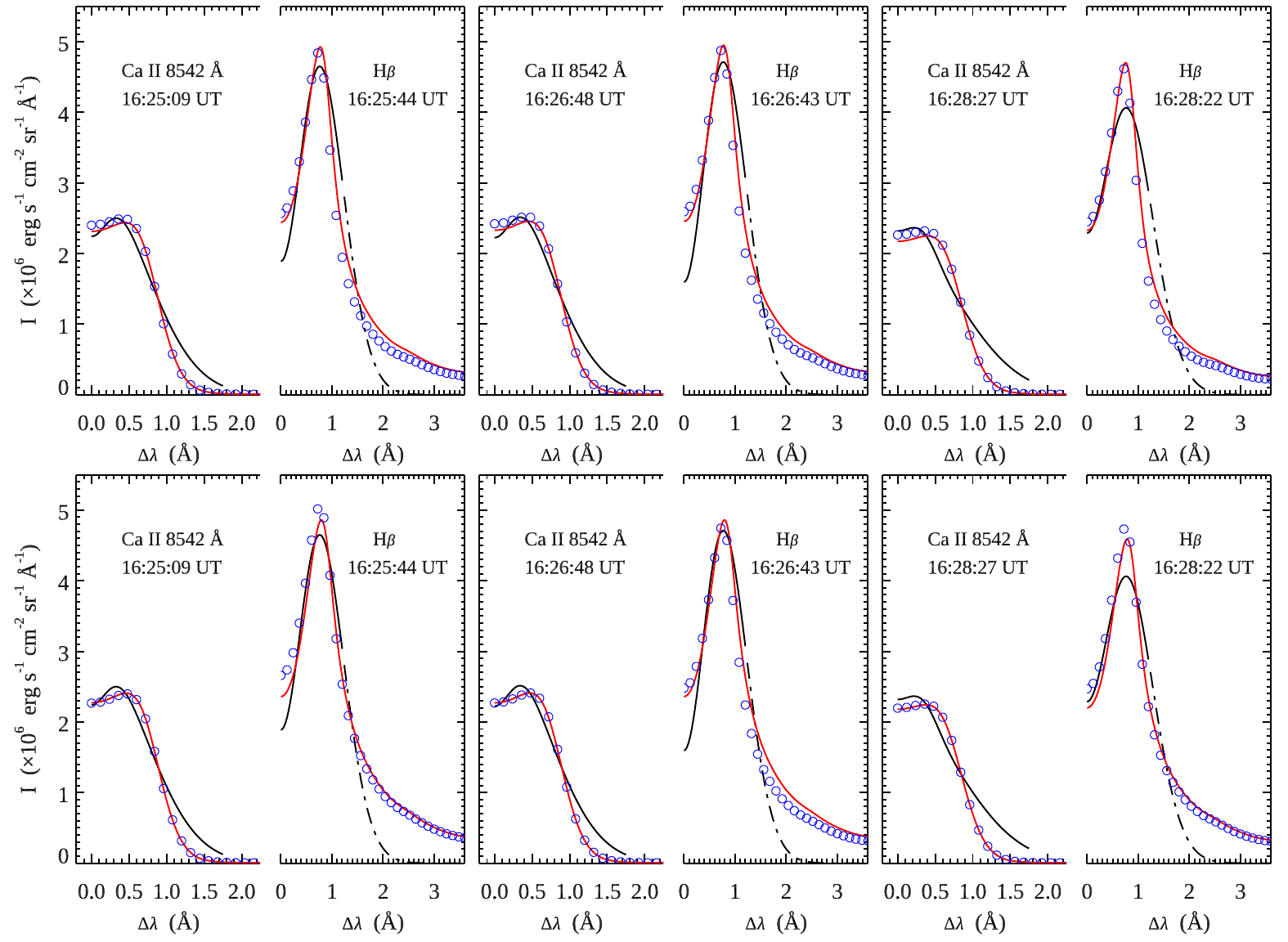}
\caption{Same as Figure~\ref{fig6}, but with fixed $\vt = 20$\,\kms.}
\label{fig7}
\end{figure}

\section{Inversion results}

Figure~\ref{fig6} displays symmetric double-Gaussian fits, as proxies
of the observed \ca\ and \hb\ line profiles, and their fits
corresponding to the model~1 and 2 in Table~\ref{tab3} that minimize
the merit function within the fine grids in Table~\ref{tab2} for the
photospheric Ca abundance. Both models roughly reproduce the basic
features of the observed profiles, such as the central reversal of the
\hb\ line, central and wing intensities of the \ca\ line. However, the
models fail to capture central reversals of \ca, prominent in the
observed profiles (Figure~\ref{fig4}: top frames).
The integrated intensities $E$ of the best-fit synthetic \ca\ profiles
and symmetric fits agree well both for the double- and single-line
inversions (Table~\ref{tab1}). However, the integrated intensities $E$
of the best-fit synthetic \hb\ profiles are always larger than the
symmetric double-Gaussian fits, whilst the synthetic \hb\ reversals
are mostly smaller than the observed reversals. These differences are
due to the Gaussian wings of the symmetric \hb\ fits
(Figure~\ref{fig4}: bottom frames), with intensities lower than the
wing intensities of the Voigt profiles that represent the synthetic
\hb\ profiles at high electron densities (Figure~\ref{fig6}), due to a
significant pressure broadening computed by the MALI code.
  
Resulting model parameters of the double-line inversion in
Table~\ref{tab3} show constancy of the plasma kinetic temperature at
8.7\,kK and decrease of gas pressure from 9.5 to 8.7\,dyn\,cm$^{-2}$,
surface electron density from $0.72 \times 10^{12}$ to $0.67 \times
10^{12}$\,\cmcub, and core electron density from $2.74 \times 10^{12}$
to $2.53 \times 10^{12}$\,\cmcub\ over the time interval of
3\,min. High electron densities are consistent with the results of
\citet{Jejcicetal2018} based on SDO/HMI continuum observations. The
geometrical slab thickness and the microturbulent velocity do not show
any trend. Reproducing the observed line profiles requires thicknesses
of $4.5 - 5.1$\,Mm and a microturbulence of 24\,\kms. The line center
optical thicknesses $\tau_0$ of the \ca\ and \hb\ lines are given in
the last column of Table~\ref{tab1} separately for the model~1 and 2
(in parentheses), for double- and single-line inversions marked by
\ddag\ and \dag, respectively. Because the synthetic \ca\ profiles
lack the central reversals ($r=1.00$) the optical thicknesses
$\tau_0($\ca) are lower limits. Similarly, the optical thicknesses
$\tau_0($\hb) are lower limits because the synthetic \hb\ reversal
depths are always smaller than the observed reversals.

However, there is a systematic difference between the resulting model
parameters yielded by the single-line inversion of \ca\ and
\hb\ profiles (compare rows in Table~\ref{tab3} starting by \dag). For
both models, fitting the \ca\ profiles requires higher temperatures of
approximately 1\,kK, smaller gas pressures, higher microturbulence,
and about two-times higher surface electron densities, and about $0.7
\times 10^{12}$\,\cmcub\ higher core electron densities than those
needed for fitting single \hb\ profiles. This suggests that the
\ca\ and \hb\ lines may originate from different parts of the
inhomogeneous loops although similar geometrical thicknesses are
required in the single-line inversions. Thus it is likely that the
model parameters yielded by the double-line inversion are weighted
averages of the parameter distributions along the LoS. Most of the
double-line model parameters fall between the values inferred by
single-line inversions.

There is some ambiguity in that models~1 and 2 differ in the values of
gas pressure and geometrical thickness, yet produce very similar
profiles as shown in Figure~\ref{fig6} by the red lines and blue
circles and also by similar $\chi2$ values in the last column of
Table~\ref{tab3}. Models~1 and 2 thus represent a range of valid
solutions, which only improved future diagnostics can improve.

To examine the effect of an increase in Ca abundance, the inversion is
repeated for the fine grids of Table~\ref{tab2}, but using synthetic
\ca\ profiles computed for the coronal Ca abundance. The resulting
best-fit parameters are given in Table~\ref{tab4} for comparison with
Table~\ref{tab3}. Although the inversions often hit the limits of
parameter ranges in Table~\ref{tab2} yielding a temperature of 10\,kK,
gas pressure of 7 and 10\,dyn\,cm$^{-2}$, and slab geometrical
thicknesses of 4.5 and 2\,Mm; the coronal Ca abundance is still
compatible with high electron number densities of about $1-2 \times
10^{12}$\,\cmcub.

  \begin{table}
    \centering
    \caption{Parameters of the best-fit model~1 and 2 (in parenthesis)
      for the double- (\ddag) and single-line (\dag) inversions taking the photospheric Ca abundance.}
    \label{tab3}
    \begin{tabular}{ c l c D@{\hspace{-2pt}}D c c c c c }
      \hline
      \hline
      \multirow{2}{*}{hh:mm:ss\,UT} & \multirow{2}{*}{Inversion} & $T$  &  \multicolumn4c{$\pg$} & $D$  & $\vt$ & $n_{\rm e}({\rm s})$ & $n_{\rm e}({\rm c})$ & $\chi^2$\\
                                 &                            & (kK) &  \multicolumn4c{(dyn\,cm$^{-2}$)} & (Mm) &    (\kms)   & ($\times 10^{12}$\,\cmcub) & ($\times 10^{12}$\,\cmcub) & ($1\sigma$) \\
      \hline  
                             & \ddag\ \ca\ + \hb\ & 8.7 (8.6) & 9.5 & (12.4) & 4.5 (2.2) & 24 (26) & 0.72 (0.83) & 2.74 (3.33)  & 2.16 (2.13) \\ 
                   16:25:09  & \dag\ \ca\         & 9.3 (9.3) & 10.4 & (14.4) & 5.9 (2.6) & 26 (28) & 1.33 (1.84) & 3.43 (4.70)  & 1.03 (0.88) \\ 
                   16:25:44  & \dag\ \hb\         & 8.4 (8.3) & 11.0 & (15.0) & 6.0 (3.0) & 20 (24) & 0.64 (0.75) & 2.68 (3.25)  & 1.08 (1.57) \\ 
      \hline
                             & \ddag\ \ca\ + \hb\ & 8.7 (8.6) & 9.1 & (11.2) & 5.1 (3.0) & 24 (24) & 0.70 (0.76) & 2.64 (3.04)  & 2.51 (2.61) \\ 
                   16:26:48  & \dag\ \ca\         & 9.3 (9.3) & 10.4 & (14.0) & 6.0 (2.8) & 26 (28) & 1.33 (1.79) & 3.43 (4.58)  & 1.07 (1.01) \\ 
                   16:26:43  & \dag\ \hb\         & 8.4 (8.3) & 11.0 & (15.0) & 6.0 (3.0) & 22 (24) & 0.64 (0.75) & 2.68 (3.25)  & 1.83 (2.58) \\ 
      \hline
                             & \ddag\ \ca\ + \hb\ & 8.7 (8.5) & 8.7 & (11.8) & 4.5 (2.0) & 26 (26) & 0.67 (0.72) & 2.53 (3.00)  & 4.15 (3.49) \\       
                   16:28:27  & \dag\ \ca\         & 9.6 (9.7) & 11.0 & (15.0) & 5.9 (3.0) & 28 (30) & 1.74 (2.56) & 3.68 (4.99)  & 2.04 (1.45) \\       
                   16:28:22  & \dag\ \hb\         & 8.3 (8.6) & 11.0 & (10.0) & 5.7 (2.0) & 22 (30) & 0.60 (0.69) & 2.49 (2.71)  & 1.37 (1.13) \\       
      \hline
    \end{tabular}
  \end{table}

  \begin{table}
    \centering
    \caption{Same as Table~\ref{tab3} for the coronal Ca abundance.}
    \label{tab4}
    \begin{tabular}{ c l D@{\hspace{-2pt}}D D@{\hspace{-2pt}}D c c c c c }
      \hline
      \hline
      \multirow{2}{*}{hh:mm:ss\,UT} & \multirow{2}{*}{Inversion} & \multicolumn4c{$T$}  &  \multicolumn4c{$\pg$}    & $D$  & $\vt$ & $n_{\rm e}({\rm s})$ & $n_{\rm e}({\rm c})$ & $\chi^2$\\
                                 &                            & \multicolumn4c{(kK)} &  \multicolumn4c{(dyn\,cm$^{-2}$)} & (Mm) &    (\kms)   & ($\times 10^{12}$\,\cmcub) & ($\times 10^{12}$\,\cmcub) & ($1\sigma$) \\
      \hline  
                             & \ddag\ \ca\ + \hb\ &  9.1 & (8.9)  &  7.0 & (10.0) & 5.6 (2.0) & 24 (26) & 0.77 (0.90) & 2.28 (3.05)  & 2.64 (3.05) \\ 
                   16:25:09  & \dag\ \ca\         & 10.0 & (9.8)  & 10.4 & (11.9) & 4.5 (2.6) & 28 (28) & 2.04 (2.12) & 3.45 (3.96)  & 0.91 (0.88) \\ 
      \hline
                             & \ddag\ \ca\ + \hb\ &  9.0 & (8.9)  &  7.0 & (10.0) & 5.4 (2.0) & 24 (26) & 0.71 (0.90) & 2.24 (3.05)  & 3.12 (3.76) \\ 
                   16:26:48  & \dag\ \ca\         & 10.0 & (9.8)  & 10.4 & (11.8) & 4.5 (2.7) & 28 (28) & 2.04 (2.10) & 3.45 (3.93)  & 1.03 (1.01) \\ 
      \hline
                             & \ddag\ \ca\ + \hb\ & 9.1  & (9.1)  &  7.0 & (10.0) & 4.5 (2.0) & 26 (28) & 0.76 (1.07) & 2.27 (3.19)  & 4.47 (6.22) \\       
                   16:28:27  & \dag\ \ca\         & 10.0 & (10.0) &  9.8 & (13.6) & 4.5 (2.0) & 28 (30) & 1.92 (2.70) & 3.25 (4.50)  & 2.14 (1.48) \\       
      \hline
    \end{tabular}
  \end{table}

\begin{figure}
\includegraphics[width=\textwidth,bb = 24 15 520 340]{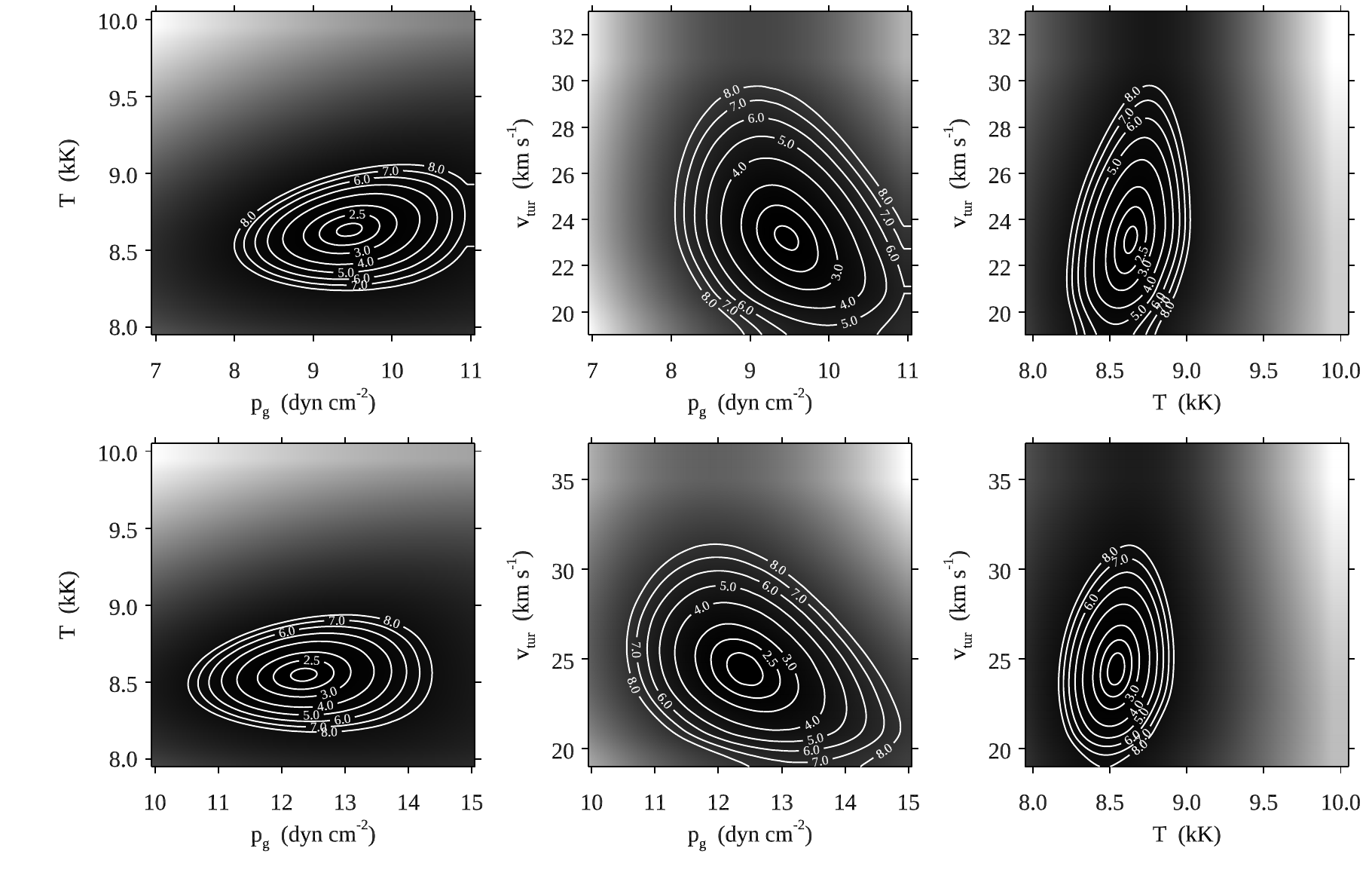}
\caption{Distribution of the $\chi^2$ function near its minima for
  double-line inversion results at $\approx$\,16:25\,UT
  (Table~\ref{tab3}: first row). Top and bottom rows correspond to the
  model~1 and 2. Pairs of complementary parameters are fixed to the
  values in the table. Darker shades reveal lower $\chi^2$ values,
  which are also outlined by selected isocontours labeled in $1\sigma$
  units. It is assumed that $1\sigma = 2 \times
  10^{-6}$\,erg\,s$^{-1}$\,cm$^{-2}$\,sr$^{-1}$\,Hz$^{-1}$.}
\label{fig8}
\end{figure}

The uniqueness and precision of the inversion within a given model
grid can be checked by exploring a distribution of $\chi^2$ values
over the parameter space. Figure\,\ref{fig8} displays three cuts
through the $\chi^2(T, \pg, D, \vt)$ hypercube out of six possible
combinations ($T - \pg$, $T - \vt$, $T - D$, $\pg - \vt$, $\pg - D$,
$\vt - D$) for the parameters in the first row of
Table~\ref{tab3}. The contour maps show the $\chi^2$ isosurfaces for
pairs of complementary parameters ($D, \vt$; $\pg, D$; etc.) fixed to
the values in Table~\ref{tab3}. They demonstrate that within fine
grids~1 and 2 there is only one optimal solution. The well-localized
and isolated minima prove that temperature, microturbulent velocity,
gas pressure, and also electron densities are well defined by the
inversion process. A comparison of the top and bottom frames of
Figure~\ref{fig8} provides an interesting insight into relations
between individual parameters. Narrowing the slab from 4.5\,Mm (top
frames) to 2.2\,Mm (bottom frames) invokes an increase of gas pressure
from 9.5 to 12.4\,dyn\,cm$^{-2}$ and also an increase of the central
electron density $\nec$ from about $2.7 \times 10^{12}$\,\cmcub\ to
$3.3 \times 10^{12}$\,\cmcub. This also holds for other double-line
inversion results (Table~\ref{tab3}). Considering a fine-thread nature
of the flare loops (which apparently consist of threads with
cross-sectional widths about 100\,km \citep{Jingetal2016}), possibly
small volume filling factor, and natural anti-correlations $\pg - D$
and $\nec - D$, then the given values of the gas pressure and electron
densities are only lower limits.

\section{Ratio $E(8542)/E($\hb) in a high-pressure regime}

\citet{GouttebrozeHeinzel2002} investigated a diagnostic potential of
the ratio of the \ca\ and \hb\ integrated intensities $E(8542)/E($\hb)
employing an extensive grid of low-pressure models up to
1\,dyn\,cm$^{-2}$, characteristic of quiescent prominences. They
demonstrated that the ratio is less than one for most of the models
(Figure~9 therein) and there exists a statistical correlation between
the ratio $E(8542)/E($\hb), the gas pressure $\pg$, and the
temperature $T$ for $\pg$ ranging from $\approx 0.1$ to
1\,dyn\,cm$^{-2}$. Inspired by these conclusions, we investigate
possible existence of such a correlation in high-pressure regimes
beyond 1\,dyn\,cm$^{-2}$ using the fine grids in Table~\ref{tab2}.

\begin{figure}
\centering
\includegraphics[width=0.49\textwidth]{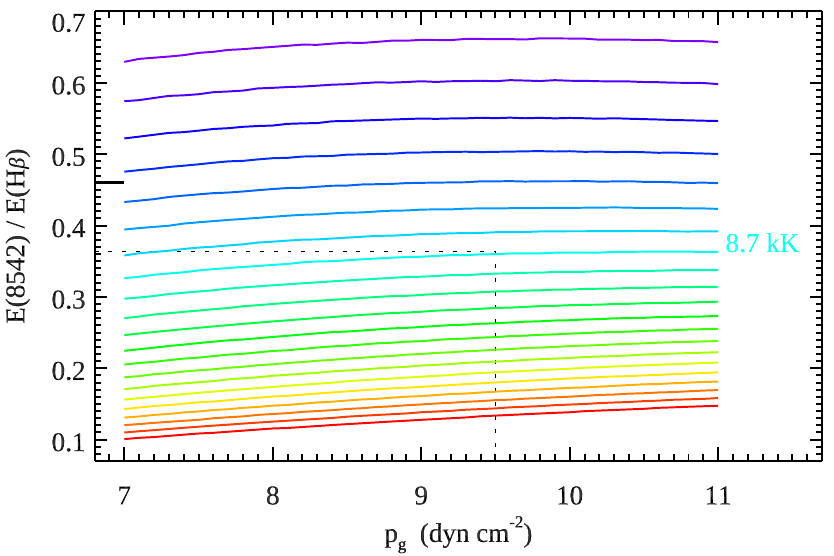}
\includegraphics[width=0.49\textwidth]{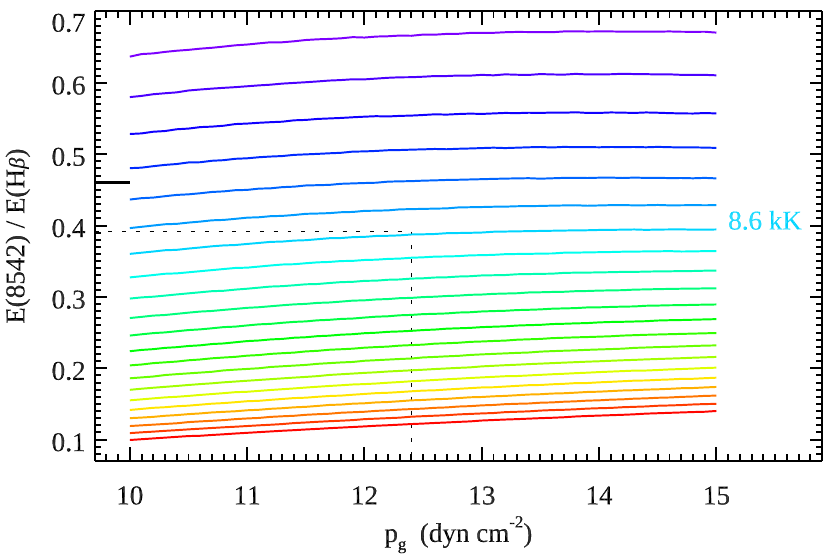}
\caption{Left frame: ratio of the integrated intensities
  $E(8542)/E($\hb) of the synthetic \cawav\ and \hb\ line profiles for
  the fine grid~1 (Table~\ref{tab2}) taking fixed $D = 4.5$\,Mm and
  $\vt = 24$\,\kms\ (Table~\ref{tab3}: $\approx$\,16:25\,UT, first
  row). The short horizontal bar indicates the observed ratio. The
  dotted lines indicated corresponding results of double-line
  inversion. Color lines correspond to temperatures ranging from
  8.0\,kK (violet) to 10.0\,kK (red) with an increment of 0.1\,kK per
  line. Right frame: same as the left frame for the fine grid~2 taking
  fixed $D = 2.2$\,Mm and $\vt = 26$\,\kms.}
\label{fig9}
\end{figure}

The left frame of Figure~\ref{fig9} illustrates the dependence
``ratio\,$-\pg-T$'' for synthetic \ca\ and \hb\ profiles computed by
the fine grid~1 taking $D = 4.5$\,Mm and $\vt = 24$\,\kms, i.\,e., for
results of the double-line inversion at $\approx$\,16:25\,UT. The
right frame shows the ratio $E(8542)/E($\hb) for the fine grid~2
taking $D = 2.2$\,Mm and $\vt = 26$\,\kms\ (Table~\ref{tab3}: first
row). The plots show that: (i) the ratio $E(8542)/E($\hb) is
constrained within the limits $0.1 - 0.7$; (ii) higher ratios
correspond to lower temperatures; and (iii) there is a slight increase
of the ratio with the gas pressure at higher temperatures. The dotted
lines show the ratios for the parameters of model~1 and 2
(Table~\ref{tab3}: first row). The ratio of corresponding symmetric
\ca\ and \hb\ fits $4.90/10.66 \approx 0.46$ (Table~\ref{tab1}) is
indicated in Figure~\ref{fig9} by the short horizontal bar. The ratio
discrepancy is due to the Gaussian wings of the symmetric \hb\ fits
(Figure~\ref{fig4}: bottom frames) with intensities which are lower
than the wing intensities of Voigt profiles representing synthetic
\hb\ profiles at high electron densities
(Figure~\ref{fig6}). Obviously, observations of full \hb\ profiles,
including near and far wings, are needed for better comparison of
synthetic and observed $E(8542)/E($\hb) ratios in the high-pressure
regime.

\newpage
\section{Discussion and conclusions}

We present \cawav\ and \hb\ imaging spectroscopy of well-developed
off-limb flare loops pertinent to the X8.2-class flare acquired about
20\,min after the flare peak over an interval of about 3\,min. We
analyze quasi-simultaneous emission line profiles averaged over small
bright patches at the top of the flare loop apex. The \ca\ and
\hb\ lines show central reversals surrounded with slightly asymmetric
peaks.
Following the absolute calibration and spatial alignment of the data,
extended grids of cool loop models are constructed, using the 1D
non-LTE code MALI. The grids are then used for the spectral
inversions.  Double- and single-line inversions are made, treating the
\ca\ and \hb\ lines together or separately.

Estimates of the kinetic temperature 8.7\,kK, the gas pressure
9.1\,dyn\,cm$^{-2}$, the microturbulent velocity 24\,\kms, and
electron number density $2.6 \times 10^{12}$\,\cmcub\ are obtained for
a geometrical thickness of 4.5\,Mm of the bright region at the top of
the flare loop arcade. These medians values gained from
  double-line inversions (Table~\ref{tab3}) are representative of the
cool component of multithermal flare loops composed of cool and hot
strands, as suggest by Hinode's EIS and SDO/AIA observations
\citep{Doscheketal2018,Jejcicetal2018,Kuridzeetal2019}. The profile
characteristics and resulting model parameters imply relatively high
electron density of the order of $10^{12}$\,\cmcub, which is very
likely a lower limit estimate due to: (i) significant fine-scale
structuring of the flare loops, revealed e.\,g. by
\citet{Jingetal2016}, with presumably small volume filling factor and
(ii) underestimation of reversal depths $r$ by our 1D
isothermal-isobaric models. Nevertheless, our inversion-based estimate
confirms high electron densities in cool loops as reported in
\citet{Jejcicetal2018} and also in a number of earlier studies
\citep{Svestka1972,Svestkaetal1987,
  HeinzelKarlicky1987,Hieietal1983,Hieietal1992}.

Another noteworthy feature of the flare loop plasma is a strong
turbulence manifested through a large non-thermal broadening of
lines. Here we find that the microturbulent velocity about
24\,\kms\ is needed to explain the profiles of \ca\ and
\hb\ lines. Our values of the microturbulent velocity agree quite well
with those obtained by \citet[see Figure~3d]{Brannon2016} and
\citet[see Figures~8, 9, and Section~5]{Mikulaetal2017} at the flare
loop apices.

There is a two-order-of-magnitude difference between electron
densities found in this study and in \citet{Jejcicetal2018} with those
inferred in \citet{Doscheketal2018} using the Hinode's EIS spectra of
hot loops. The latter study reports a conservative density of the
order of $10^{10}$\,\cmcub\ in the hot loops within the same
multithermal flare loop system. Remarkably, it also shows a coronal Ca
abundance at the cusp height $\approx 42$\,Mm, which steadily changes
from coronal to photospheric towards the solar limb. Figure~4 therein
suggests an intensity ratio Ca\,{\sc xiv}/Ar\,{\sc xiv} of about $2-3$
indicating transformative coronal-to-photospheric Ca abundance for the
bright patches at the cool loop apex at height $\approx 17$\,Mm. Our
inversions show that high electron number densities of about $1-2
\times 10^{12}$\,\cmcub\ are consistent with the upper limit of
coronal Ca abundance.

Interpreting the \ca\ and \hb\ line profiles (Figures~\ref{fig4} and
\ref{fig6}) requires a median optical thickness of 4.5 for the former
and $\approx 70$ or $\approx 44$ for the latter, depending on the
adopted geometrical thickness. Due to the unsatisfactory fits of
central reversals of both lines (i.\,e. absent, or not deep enough)
these values are very likely lower limits.

To explain the absence of \ca\ reversals, we constrained the fine
grids~1 and 2 with a lower microturbulent velocity of 20\,\kms, and
photospheric Ca abundance. The inversion yields the best fits shown in
Figure~\ref{fig7} with the \ca\ profiles having central reversal that
are very similar to the observed. However, the overall \ca\ fit of the
profile wings is less satisfactory than if the whole range of
microturbulent velocities is considered (Figure~\ref{fig6}). Fitting
the \ca\ reversal requires the following best-fit parameters for the
fine grid~1 (grid~2): $T= 8.6~(8.5)$\,kK, $\pg =
9~(11.9)$\,dyn\,cm$^{-2}$, $D = 6~(3)$\,Mm, and $\nec =
2.5~(3.04)$\,\cmcub\ corresponding to $\chi^2 = 4.48~(6.05)$ and
$\tau_0($\ca) = 6.9 (6.4). We interpret these results as a signature
of insufficient pressure broadening in the wings of the synthetic
\ca\ profiles that is compensated by an increased microturbulence. The
latter and the instrumental broadening remove the \ca\ reversal but
yield a better overall fit. High electron densities produce enhanced
and extended pressure-broadened \hb\ profiles (Figures~\ref{fig6} and
\ref{fig7}). For \ca\ lines we used the Stark and van der Waals
broadening parameters suggested by \citet{ShineLinsky1974} but this
ideally should be based on more modern damping parameters.

Our future study will repeat the density diagnostics of cool flare
loops employing more sophisticated modeling of the whole loops using
the 2D non-LTE radiative transfer code MALI2D
\citep{HeinzelAnzer2001,Heinzeletal2005,Gunaretal2007}. The current
study will provide initial estimates of the plasma parameters. This
should give a more satisfactory coincidence of the observed and
synthetic profiles, a better fit to the central reversals, and a more
realistic reproduction of the peak asymmetries of the \ca\ and
\hb\ line profiles through including bulk motions along the
loops. Such results will also improve understanding of the importance
of flare loops on other flaring stars, particularly those that produce
superflares \citet{HeinzelShibata2018}.

\begin{acknowledgements}
This research has received funding by the S\^{e}r Cymru~II,
part-funded by the European Regional Development Fund through the
Welsh Government. HM is partly funded by an STFC consolidated grant to
Aberystwyth University. The Swedish 1-m Solar Telescope is operated on
the island of La Palma by the Institute for Solar Physics of Stockholm
University in the Spanish Observatorio del Roque de los Muchachos of
the Instituto de Astrof\'{\i}sica de Canarias. The Institute for Solar
Physics is supported by a grant for research infrastructures of
national importance from the Swedish Research Council (registration
number 2017-00625). The work of DK was supported by Georgian Shota
Rustaveli National Science Foundation project FR17\_323. JK
acknowledges the project VEGA 2/0004/16. This work was also supported
by the project ITMS No. 26220120029, based on the operational Research
and Development Program financed from the European Regional
Development Fund. The authors acknowledge support from the grant
19-17102S of the Czech Funding Agency. SJ acknowledges the financial
support from the Slovenian Research Agency No. P1-0188. The data were
acquired within Spanish SST time allocation. This research has made
use of NASA's Astrophysics Data System. The authors wishes to thank
the anonymous referee who provided constructive remarks that greatly
improved the quality of this paper.
\end{acknowledgements}

\facility{SST(CRISP,CHROMIS)}

\end{document}